\documentclass[]{jfm}

\usepackage{graphicx}
\usepackage{epstopdf,epsfig}

\newcommand{\bm}[1]{\boldsymbol{#1}}

\shorttitle{
Energy-flux vector in anisotropic turbulence
}
\title{
Energy-flux vector in anisotropic turbulence:
application to rotating turbulence
}

\shortauthor{N. Yokoyama and M. Takaoka}

\author{Naoto Yokoyama\aff{1}
\corresp{\email{n.yokoyama@mail.dendai.ac.jp}},
\aunote{Present address: Department of Mechanical Engineering, Tokyo Denki University, Adachi, Tokyo, 120-8551, Japan.}
\and
Masanori Takaoka\aff{2}
}

\affiliation{\aff{1}Department of Mechanical Science and Bioengineering, Osaka University, Toyonaka 560-8531, Japan
\aff{2}Department of Mechanical Engineering, Doshisha University, Kyotanabe 610-0394, Japan}

\begin{document}

\maketitle

\begin{abstract}
Energy flux plays a key role in the analyses of energy-cascading turbulence.
In isotropic turbulence, the flux is given by a scalar as a function of the magnitude of the wavenumber.
On the other hand, the flux in anisotropic turbulence should be a geometric vector that has a direction as well as the magnitude, and depends not only on the magnitude of the wavenumber but also on its direction.
The energy-flux vector in the anisotropic turbulence cannot be uniquely determined in a way used for the isotropic flux.
In this work, introducing two ansatzes, net locality and efficiency of the nonlinear energy transfer, we propose a way to determine the energy-flux vector in anisotropic turbulence by using the Moore--Penrose inverse.
The energy-flux vector in strongly rotating turbulence is demonstrated based on the energy transfer rate obtained by direct numerical simulations.
It is found that the direction of the energy-flux vector is consistent with the prediction of the weak turbulence theory in the wavenumber range dominated by the inertial waves.
However, the energy flux along the critical wavenumbers predicted by the critical balance in the buffer range between in the weak turbulence range and the isotropic Kolmogorov turbulence range is not observed in the present simulations.
This discrepancy between the critical balance and the present numerical results is discussed and the dissipation is found to play an important role in the energy flux in the buffer range.
\end{abstract}

\section{Introduction}

It is one of the most important subjects in turbulence research
how energy,
which is provided by external force and dissipated by viscosity,
is transferred among scales via nonlinear interactions.
In the Kolmogorov turbulence,
the energy cascades from large-scale eddies to the small-scale ones
via nonlinear interactions.
In the weak-wave turbulence,
the energy is transferred via nonlinear {\em resonant\/} interactions among waves.

The assumption of the weak nonlinearity
that the linear time scale is much smaller than the nonlinear time scale
is violated at small or large wavenumbers
in almost all the wave turbulence systems~\citep{Biven200128,Biven200398,newell01}.
In this case,
the weak-wave turbulence and the strong turbulence
coexist~\citep{PhysRevX.8.031066,Vinen2002,PhysRevE.89.012909}.
According to the conjecture of the critical balance~\citep{goldreich1995toward,nazarenkobook,nazarenko3488critical},
the energy is considered to be transferred
along the wavenumbers
at which the linear wave period is comparable with the eddy turnover time
of the isotropic Kolmogorov turbulence
in the buffer range between the wavenumber ranges
of the weak-wave turbulence and the isotropic Kolmogorov turbulence.
The conjecture is being eagerly tested by using numerical simulations~\citep{PhysRevLett.110.145002,PhysRevLett.116.105002,PhysRevX.8.031066,doi:10.1063/1.3693974}.

The scale-by-scale energy cascade is often investigated by energy flux.
The constancy of the energy flux in the wavenumber space
is intensively examined as a corollary of the energy cascade
in the homogeneous isotropic turbulence (HIT).
While the Kolmogorov theory predicts scaling properties,
the quantitative feature of the flux for sufficiently large Reynolds numbers
has been examined theoretically and numerically.

However, the energy flux in anisotropic turbulence is less elucidated
because the analytical expression for the flux is not known
as contrasted in HIT.
While the flux can be treated as a scalar in HIT,
it should be treated as a vector in anisotropic turbulence
even if it is homogeneous in the real space.
In this paper,
the definition of the energy-flux vector is proposed,
and it is applied to rotating turbulence.

\begin{figure}
 \centerline{
   \includegraphics[scale=1.2]{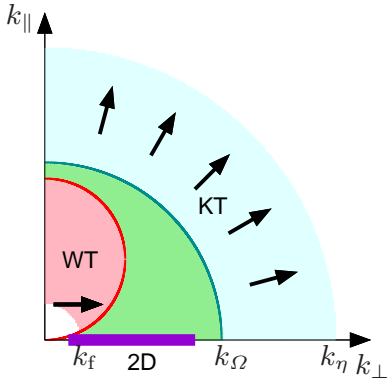}
 }
 \caption{
 Schematic energy flux in rotating turbulence.
 The wavenumbers of the two-dimensional vortex are indicated by the thick purple line
 on the $k_{\perp}$ axis.
 The wavenumber ranges of the weak-wave turbulence,
 the buffer range and the isotropic Kolmogorov turbulence
 are respectively coloured red, green and blue.
 }
 \label{fig:coexistrt}
\end{figure}

Let us consider a homogeneous anisotropic turbulence system
which has one distinguishing direction, say the $z$ direction,
and is statistically isotropic in the $x$ and $y$ directions perpendicular to the $z$ direction.
It is convenient to investigate the energy transfers
in the $k_{\perp}$-$k_{\|}$ plane,
where $k_{\perp} = |\bm{k}_{\perp}| = (k_x^2+k_y^2)^{1/2}$ and $k_{\|} = |k_z|$.
In figure~\ref{fig:coexistrt},
the energy flux expected in the rotating turbulence is schematically drawn.
The rotating turbulence
is a typical turbulence system
where
the weak-wave turbulence of inertial waves and the isotropic Kolmogorov turbulence of eddies
as well as the two-dimensional columnar vortex
coexist~\citep{:/content/aip/journal/pof2/26/3/10.1063/1.4868280,PhysRevFluids.2.092602}.
In the wavenumber range
where the linear period of the inertial wave is shorter than the eddy turnover time,
the weak turbulence of the inertial waves is dominant
(WT in figure~\ref{fig:coexistrt}).
According to the weak turbulence theory,
the resonant interactions among the inertial waves
transfer energy to waves which have small scales perpendicular to the rotational axis
without changing its scales parallel to the rotational axis~\citep{FLM:465066,PhysRevE.68.015301,Yarom2014}.
On the other hand,
the isotropic energy transfer due to the isotropic Kolmogorov turbulence appears
in the wavenumber range
where the Coriolis period is longer than the eddy turnover time~\citep{mininni_rosenberg_pouquet_2012},
i.e., the wavenumber is larger than the Zeman wavenumber $k_{\Omega}$ (KT in figure~\ref{fig:coexistrt}).
However,
there is no concrete theory that quantitatively gives the energy transfer in the buffer range (green in figure~\ref{fig:coexistrt}).
The critical balance predicts the isotropisation due to the redistribution
of energy that is transferred anisotropically to the buffer range.
If the local energy-flux vectors can be obtained,
the arrows of the energy flux in the buffer range are added to figure~\ref{fig:coexistrt}.

The energy flux in HIT is defined as
the flux going through the sphere with a radius $|\bm{k}| = k$.
To evaluate the energy flux in the anisotropic turbulence,
the fluxes going through the cylindrical surface with a radius $|\bm{k}_{\perp}| = k_{\perp}$
and through the planes with $|k_z| = k_{\|}$
have been used as a simple tool
in anisotropic turbulence~\citep{PhysRevE.90.023005,FLM:409533,doi:10.1175/JPO3027.1}.
The conical energy flux,
which evaluates the flux going through the surface of a cone in the wavenumber space,
was also proposed by \citet{doi:10.1063/1.5109856}.
The perpendicular energy flux, for example,
is obtained by integration over all the parallel wavenumbers,
and it corresponds to the flux going through a line parallel to the parallel-wavenumber axis
in figure~\ref{fig:coexistrt}.
These integrated energy fluxes inevitably consist of
contributions from multiple turbulence ranges,
the weak-wave turbulence, the buffer range,
and the isotropic Kolmogorov turbulence.
Thus, the local energy flux is expected to be identified
to see the paths of the energy transfer.

The locality of the net energy transfer in the wavenumber space is required
for the concept of the {\em flux\/} to be reasonable.
The locality is one of the major concept naturally assumed in the cascading theory,
originating from Richardson's poetical note~\citep{richardson_lynch_2007}
and sophisticated by Kolmogorov theory (K41)~\citep{K41a}.
On the other hand,
the locality as well as the correction of some aspects of K41
has been widely studied in the literature,
starting with the pioneering work of \citet{kraichnan_1959,kraichnan_1971}.
It should be noted that the nonlocal energy transfer should be distinguished
from the energy transfer in a nonlocal triad that is a flat triad~\citep{doi:10.1063/1.858309,:/content/aip/journal/pofa/5/3/10.1063/1.858651}.
The sweeping effect,
in which small-scale eddies are advected by a large-scale eddy,
does not make the net nonlocal energy transfer~\citep{doi:10.1063/1.858296}.
In fact,
\citet{doi:10.1063/1.5109856} reported that
the net energy transfer is mainly local in the rotating turbulence
when the triad interactions are considered as one-to-one interactions.
According to the net locality,
diffusion models given by partial differential equations in the wavenumber space
are often employed in the weak turbulence~\citep{DYACHENKO199296,GALTIER201984,doi:10.1175/1520-0485(1985)015<1378:CAPOTN>2.0.CO;2,zak_book}
and in the Kolmogorov turbulence~\citep{PhysRevLett.92.044501,doi:10.1063/1.1762300}.
The diffusion models are also applied to anisotropic turbulence systems~\citep{Galtier_2010,PhysRevE.79.035401}.

In this paper,
the energy-flux vector is obtained by using the Moore--Penrose inverse,
to identify the direction of the energy flux in the anisotropic turbulence.
The organisation of this paper is as follows.
The procedure to obtain the energy-flux vector is proposed,
and it is verified by examining the energy-flux vector in HIT
in \S\ref{sec:formulation}.
The energy-flux vectors in the rotating turbulence
 are presented
in \S\ref{sec:application}.
The validity of the proposed vector in anisotropic turbulence
is examined by comparing it with the weak turbulence theory.
The energy flux for a more commonly-used external forcing is shown in detail.
It is consistent with the weak turbulence theory
in the wavenumber range where the energy spectrum agrees with the weak turbulence theory,
but is observed to be inconsistent with that predicted by the critical balance in the buffer range in this simulation.
In \S\ref{sec:discussion},
the reason for the inconsistency
between the fluxes in the present simulation and in the critical balance
is clarified.
The last section is devoted to the summary.

\section{Formulation of energy-flux vector}
\label{sec:formulation}

In this section,
the energy-flux vector is proposed by introducing two ansatzes.
The procedure to numerically obtain the energy-flux vector is described with a concrete example,
which is the rotating turbulence.
The rotating turbulence is a typical homogeneous anisotropic turbulence system
that has been extensively investigated,
and can be an appropriate testbed to examine the proposed idea,
since it contains different kinds of turbulence (see figure~\ref{fig:coexistrt}).
In addition,
it is easy to examine the direction of the energy-flux vectors
in the range of the inertial-wave turbulence.

\subsection{Integrated energy fluxes}
\label{ssec:integratedenergyflux}

The governing equations for the velocity $\bm{u}$ of rotating turbulence
in incompressible fluid
are the Navier--Stokes equation with the Coriolis term
and the divergence-free condition:
\begin{subeqnarray}
&
\displaystyle
  \frac{\partial \bm{u}}{\partial t}
  + (\bm{u} \bcdot \bnabla) \bm{u}
  + 2 \bm{\Omega} \times \bm{u}
  =
  - \bnabla p
  + \nu \nabla^2 \bm{u}
  + \bm{f}
  ,
  \\
  &
  \bnabla \bcdot \bm{u} = 0
 ,
 \label{eq:governingeq4rt}%
\end{subeqnarray}%
where the centrifugal force is included in the pressure $p$.
The rotation vector $\bm{\Omega}=\Omega\bm{e}_z$ is assumed to be constant.
The governing equations of the isotropic turbulence
are the same as (\ref{eq:governingeq4rt}) but $\bm{\Omega} = \bm{0}$.
The kinematic viscosity is expressed by $\nu$.
The external force $\bm{f}$ is Gaussian white.
Under the periodic boundary condition,
the governing equation~(\ref{eq:governingeq4rt}) is rewritten
in the wavenumber space as
\begin{subeqnarray}
 \displaystyle
  \frac{\partial \bm{u}_{\bm{k}}}{\partial t}
 &
 =
\displaystyle
-  \left(
 \mathsfbi{I} - \frac{\bm{k} \otimes \bm{k}}{k^2}
 \right)
 \bcdot
\left(
2 \bm{\Omega} \times \bm{u}_{\bm{k}}
+ \mathrm{i}
 \sum_{\bm{k}_1+\bm{k}_2=\bm{k}}
 (\bm{u}_{\bm{k}_1} \bcdot \bm{k}_2) \bm{u}_{\bm{k}_2}
\right)
 - \nu k^2 \bm{u}_{\bm{k}}
 + \bm{f}_{\bm{k}}
,
 \\
 &
 \bm{k} \bcdot \bm{u}_{\bm{k}} = 0
 .
\label{eq:governingeqrtw}
\end{subeqnarray}

Energy is transferred among wavenumbers via the nonlinear interactions
which come from the advection term in (\ref{eq:governingeqrtw}\textit{a}).
The energy transfer rate for a wavenumber $\bm{k}$
via the nonlinear interactions among three-wavenumber modes is symbolically written as
\begin{equation}
T_{\bm{k}} =
\left\langle
 \left.
  \frac{\partial E_{\bm{k}}}{\partial t}
  \right|_{\mathrm{NL}}
\right\rangle
 = \sum_{\bm{k}_1,\bm{k}_2} \mathcal{T}(\bm{u}_{\bm{k}}; \bm{u}_{\bm{k}_1}, \bm{u}_{\bm{k}_2})
,
\end{equation}
where $\bm{u}_{\bm{k}}$ and $E_{\bm{k}} = |\bm{u}_{\bm{k}}|^2/2$ are respectively the velocity and the energy of the wavenumber $\bm{k}$,
and $\langle \cdot \rangle$ represents the ensemble average.
Note that the energy transfer rate $T_{\bm{k}}$ is statistically equal
to the difference between energy input by the external force
and the energy dissipation rate
in the statistically steady states.
The triad-interaction function
$\mathcal{T}(\bm{u}_{\bm{k}}; \bm{u}_{\bm{k}_1}, \bm{u}_{\bm{k}_2})$ quantifies the energy transfer from or to the wavenumber $\bm{k}$
via the triad $\bm{k}+\bm{k}_1+\bm{k}_2=\bm{0}$,
and
\begin{equation}
 \mathcal{T}(\bm{u}_{\bm{k}}; \bm{u}_{\bm{k}_1}, \bm{u}_{\bm{k}_2}) =
 -\frac{\mathrm{i}}{4}
 \left\langle
 \left(\bm{k}_1 \bcdot \bm{u}_{\bm{k}_2} \right)
 \left(\bm{u}_{\bm{k}} \bcdot \bm{u}_{\bm{k}_1}\right)
 \right\rangle
 \delta_{\bm{k} + \bm{k}_1 + \bm{k}_2}
 + \mathrm{c.c.}
 + (1 \leftrightarrow 2)
,
\label{eq:triadintfunc}
\end{equation}
in both isotropic turbulence and rotating turbulence studied in this paper.
Here, $(1 \leftrightarrow 2)$ represents the terms
with the suffices $1$ and $2$ interchanged in the preceding ones.
The triad-interaction function $\mathcal{T}(\bm{u}_{\bm{k}}; \bm{u}_{\bm{k}_1}, \bm{u}_{\bm{k}_2})$
is symmetric under the interchange of $\bm{k}_1$ and $\bm{k}_2$,
and the energy detailed balance
 $\mathcal{T}(\bm{u}_{\bm{k}}; \bm{u}_{\bm{k}_1}, \bm{u}_{\bm{k}_2})
 + \mathcal{T}(\bm{u}_{\bm{k}_1}; \bm{u}_{\bm{k}_2}, \bm{u}_{\bm{k}})
 + \mathcal{T}(\bm{u}_{\bm{k}_2}; \bm{u}_{\bm{k}}, \bm{u}_{\bm{k}_1}) = 0$
 holds.
The triad-interaction function
$\mathcal{T}(\bm{u}_{\bm{k}}; \bm{u}_{\bm{k}_1}, \bm{u}_{\bm{k}_2})$
is the sum of the energy transfer
between $\bm{k}$ and $\bm{k}_1$ and that between $\bm{k}$ and $\bm{k}_2$.

To confirm the cascade theory in the homogeneous isotropic turbulence (HIT),
the energy flux
\begin{eqnarray}
 P(k) = -\int_0^k \mathrm{d}k^{\prime} T(k^{\prime})
 \label{eq:fluxiso}
\end{eqnarray}
is usually examined.
 The isotropic energy transfer rate $T(k)$
 is obtained from $T_{\bm{k}}$ by integration over the solid angle of $\bm{k}$,
 and is assumed to be a continuous function of $k=|\bm{k}|$.

Anisotropic turbulence generally has a distinguishing direction,
which is, for example, the direction of the rotational axis in the rotating turbulence.
The $z$ direction is set to be such distinguishing direction
in this paper.
The $x$ and $y$ directions are perpendicular to the distinguishing direction.
Suppose that the system statistically has the azimuthal symmetry with respect to the $z$ direction,
i.e., the azimuthal isotropy in the $x$-$y$ plane.
Then, the statistical quantities can be described in the $k_{\perp}$-$k_{\|}$ plane,
where $k_{\perp}=(k_x^2+k_y^2)^{1/2}$ and $k_{\|}=|k_z|$.

As a natural extension of the isotropic energy flux~(\ref{eq:fluxiso}) to azimuthally symmetric turbulence,
the energy fluxes perpendicular and parallel to the system's distinguishing direction
\begin{subeqnarray}
 P_{\perp}(k_{\perp}) &=
\displaystyle
 - \int_0^{k_{\perp}} \mathrm{d} k_{\perp}^{\prime}
 \int_0^{\infty} \mathrm{d} k_{\|}^{\prime}
 T(k_{\perp}^{\prime}, k_{\|}^{\prime})
 ,
\\
 P_{\|}(k_{\|}) &=
\displaystyle
- \int_0^{k_{\|}} \mathrm{d} k_{\|}^{\prime}
\int_0^{\infty} \mathrm{d} k_{\perp}^{\prime}
T(k_{\perp}^{\prime}, k_{\|}^{\prime})
\label{eq:flux1d}
\end{subeqnarray}
have been used~\citep[e.g.,][]{PhysRevE.76.056313}.
The anisotropic energy transfer rate $T(k_{\perp}, k_{\|})$
is obtained from $T_{\bm{k}}$ by integration over the azimuthal angle of $\bm{k}_{\perp}$ and the sign of $k_z$.
Because $P_{\perp}(k_{\perp})$ and $P_{\|}(k_{\|})$
are respectively obtained by the integration over $k_{\perp}$ and $k_{\|}$,
detailed local structures such as critical balance in the wavenumber space cannot be captured directly by these integrated energy fluxes.
These energy fluxes, (\ref{eq:fluxiso}) and (\ref{eq:flux1d}),
are referred to as integrated fluxes in this paper.

\subsection{Minimal-norm energy-flux vector}
\label{ssec:vectors}

To quantitatively investigate the energy-transfer mechanism in anisotropic turbulence,
the detailed structure local in the wavenumber space of the energy flux is needed to be investigated.
In this paper,
the scalar-valued energy flux defined in HIT
is extended to a vector-valued energy flux in the anisotropic turbulence.
Because of the energy cascade,
the definition of the energy flux in HIT~(\ref{eq:fluxiso})
implicitly assumes the net locality of the nonlinear interactions
and the local energy conservation in the wavenumber space:
\begin{eqnarray}
 T(k) + \frac{\mathrm{d} P(k)}{\mathrm{d} k} = 0
 .
\end{eqnarray}
As an extension to the energy flux in the anisotropic turbulence,
we assume the local energy conservation:
\begin{eqnarray}
 T_{\bm{k}} + \mathrm{div}_{\bm{k}} \bm{P}_{\bm{k}} = 0
 ,
 \label{eq:localenergyconservation}%
\end{eqnarray}%
where
$\mathrm{div}_{\bm{k}}$ is the divergence operator in the wavenumber space,
and
$\bm{P}_{\bm{k}} = (P_{x \bm{k}}, P_{y \bm{k}}, P_{z \bm{k}})$ is the energy-flux vector.

In general, the nonlinear interactions due to the advection term
contain both local and nonlocal interactions in the wavenumber space.
However, when we try to draw the {\em flux\/} as a vector field,
we must implicitly consider the flux to represent the local interactions.
Therefore,
the ansatz of the local energy conservation is a natural consequence from
the present purpose to find the energy-flux vector.
The local energy conservation~(\ref{eq:localenergyconservation}) can be
interpreted as an alternative expression of the net local interactions
to the diffusion models
where the energy transfer in the wavenumber space is approximated
by partial differential equations~\citep{Galtier_2010,PhysRevE.79.035401}.

The origin of the anisotropy of the energy flux should be described here
based on (\ref{eq:localenergyconservation}).
The anisotropy of the energy-flux vector comes from that of the energy transfer rate.
In the statistically steady state,
the anisotropy of the energy transfer rate can be explained
in terms of the energy input due to the external force
and the dissipation rate due to the viscous term.
The velocity and hence the energy can be statistically anisotropic
owing to the system's anisotropy.
In the rotating turbulence,
the anisotropy of the energy is remarkable at the small wavenumbers.
Thus, the energy input
given by the inner product of the velocity and the external force
is anisotropic owing to the anisotropy of the velocity
even if the external force is isotropic.
The dissipation rate at small wave numbers is also anisotropic
owing to the anisotropy of the energy
even if the viscous term and the dissipation rate at the large wavenumbers outside of the isotropic Kolmogorov turbulence range are isotropic.
The energy-flux vector is determined by the energy transfer rate
all over the wavenumber domain.
It is similar to the pressure obtained by solving the Poisson equation
in the real space.
Therefore, the energy-flux vector is anisotropic in the anisotropic turbulence
even if the external force and the viscous term are isotropic.

\begin{figure}
\centerline{
 \includegraphics[scale=.3]{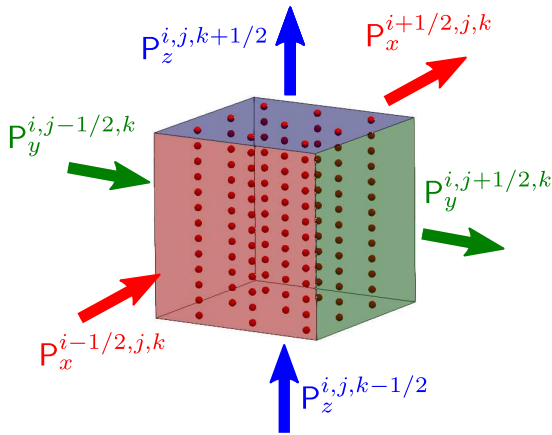}
}
 \caption{
 Coarse-graining in the wavenumber space and local energy balance.
 The red points represent the wavenumbers
 at which the energy transfer rates are evaluated.
}
 \label{fig:discretization}
\end{figure}
The local energy conservation~(\ref{eq:localenergyconservation}) is extended to a discrete formulation
that is convenient in the numerical analysis.
Let us consider the energy balance in a coarse-grained cell
that has the side lengths $\Delta k_x$, $\Delta k_y$ and $\Delta k_z$
in the wavenumber space as shown in figure~\ref{fig:discretization}.
The energy balance in the $(i,j,k)$ cell
is obtained from the local energy conservation~(\ref{eq:localenergyconservation})
as
\begin{equation}
 \mathsf{P}_x^{i+1/2,j,k} - \mathsf{P}_x^{i-1/2,j,k}
+
 \mathsf{P}_y^{i,j+1/2,k} - \mathsf{P}_y^{i,j-1/2,k}
+
 \mathsf{P}_z^{i,j,k+1/2} - \mathsf{P}_z^{i,j,k-1/2}
 = - \mathsf{T}^{i,j,k}
.
\label{eq:dp-te}
\end{equation}
The energy fluxes incoming to and outgoing from the cell,
$\mathsf{P}_x^{i \pm 1/2,j,k}$,
$\mathsf{P}_y^{i,j \pm 1/2,k}$ and
$\mathsf{P}_z^{i,j,k \pm 1/2}$,
are defined on the cell faces,
and the energy transfer rate of the cell
$\mathsf{T}^{i,j,k}$
is given as the sum of the energy transfer rates of the wavenumbers in the cell.
It is represented in a matrix-vector form as
\begin{eqnarray}
 \mathsfbi{D} \, \mathsf{P} = -\mathsf{T}
 ,
 \label{eq:dp-t}
\end{eqnarray}
where
$\mathsfbi{D} \in \mathbb{R}^{N_x N_y N_z \times (3 N_x N_y N_z - N_x N_y - N_y  N_z - N_z N_x)}$
corresponds to the difference in (\ref{eq:dp-te})
derived from the divergence operator in (\ref{eq:localenergyconservation}),
$\mathsf{P} \in \mathbb{R}^{(3 N_x N_y N_z - N_x N_y - N_y N_z - N_z N_x) \times 1}$
is a solution column vector of the flux,
and
$\mathsf{T} \in \mathbb{R}^{N_x N_y N_z \times 1}$
is the column vector of the transfer rates.
The numbers of the coarse-grained cells in the $x$, $y$ and $z$ directions
are respectively $N_x$, $N_y$ and $N_z$.
The number of the components of $\mathsf{P}$
is approximately three times larger than that of $\mathsf{T}$,
since $\bm{P}_{\bm{k}}$ consists of three components in the three-dimensional (3D) turbulence
while $T_{\bm{k}}$ is a scalar.
Note that the number of the components of $\mathsf{P}$ is smaller
than three times that of $\mathsf{T}$,
because the energy flux to or from the outer range of the computational domain does not exist.

Obviously,
the divergence matrix $\mathsfbi{D}$ has linearly independent rows.
Thus,
the solution of (\ref{eq:dp-t}) is not unique.
The nature often adopts the most efficient way
under constraints.
For example,
the minimal energy state
where the Euclidean norm of the velocity vector is minimal
is realised
under the condition that the vorticity invariants are conserved
in Euler flows~\citep{vallis_carnevale_young_1989}.
The minimal-norm flux proposed here has the minimal Euclidean norm
under the condition that the energy transfer rate is provided by the nonlinear interactions.
The selection of the minimal-norm vector is the least-action principle in this system
where the Euclidean norm of the energy-flux vector is considered as an action.
We here introduce this principle to uniquely determine energy-flux vectors,
the solution of (\ref{eq:dp-t}).

It is the Moore--Penrose inverse, which is a generalised inverse,
that can find an appropriate flux vector among the infinite number of the solutions.
For $\mathsfbi{D}$ having linearly independent rows,
the Moore--Penrose inverse is defined as
$\mathsfbi{D}^+ = \mathsfbi{D}^T (\mathsfbi{D}\mathsfbi{D}^T)^{-1}$.
The Moore--Penrose inverse selects the solution
as
$\mathsf{P}_{\ast} = - \mathsfbi{D}^+ \mathsf{T}
= \mathrm{arg min} \|\mathsf{P}\|_2$
such that (\ref{eq:dp-t}) holds.
Namely,
$\mathsf{P}_{\ast}$ selected by the Moore--Penrose inverse has the minimal Euclidean norm
among the infinite number of the solutions of (\ref{eq:dp-t}).

In addition,
the minimal-norm solution $\mathsf{P}_{\ast}$ is irrotational.
Because the divergence of the rotation is $0$
and because every null-space component of the matrix having linearly independent rows
is orthogonal to any of the row-space component,
$\mathsf{P}_{\ast}$ does not have the null-space solenoidal component.
Thus,
the use of the Moore--Penrose inverse to (\ref{eq:dp-t}) is equivalent to the assumption
that the energy-flux vector is irrotational.
The selection of the minimal-norm flux
can be interpreted that the energy transfer is ``efficient''
in the sense that the minimal-norm flux excludes local circulations of the energy transfer.

The minimal-norm solution $\mathsf{P}_{\ast}$ is obtained
by numerically solving (\ref{eq:dp-t})
as follows.
The energy transfer rate for each wavenumber mode
$T_{\bm{k}}$ in the statistically steady state
is obtained in the direct numerical simulation (DNS).
The column vector of the transfer rates $\mathsf{T}$
is composed
by coarse-graining of $T_{\bm{k}}$.
By using the generalised minimal residual (GMRES) method,
$(\mathsfbi{D}\mathsfbi{D}^T)^{-1} \mathsf{T}$ is obtained.
For the convergence criterion,
the relative error is below $10^{-10}$.
The minimal-norm vector $\mathsf{P}_{\ast}$
is obtained by applying $-\mathsfbi{D}^T$
to the vector obtained in the previous step.
\begin{figure}
\centerline{
\includegraphics[scale=.3]{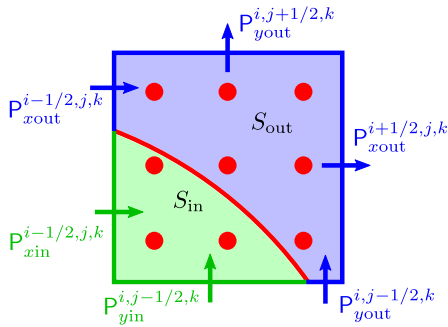}
}
\caption{
 Perpendicular energy flux through an arc.}
\label{fig:3d22d}
\end{figure}
Once $\mathsf{P}_{\ast}$ is obtained,
the energy-flux vector of the wavenumber $\bm{k}$ located at the center of the $(i,j,k)$ cell
is given as
\begin{equation}
\bm{P}_{\bm{k}} = \left(
\frac{\mathsf{P}_{\ast}^{i+1/2,j,k}+\mathsf{P}_{\ast}^{i-1/2,j,k}}{2 \Delta k_y \Delta k_z},
\frac{\mathsf{P}_{\ast}^{i,j+1/2,k}+\mathsf{P}_{\ast}^{i,j-1/2,k}}{2 \Delta k_z \Delta k_x},
\frac{\mathsf{P}_{\ast}^{i,j,k+1/2}+\mathsf{P}_{\ast}^{i,j,k-1/2}}{2 \Delta k_x \Delta k_y}
\right)
.
\end{equation}

The energy-flux vector in the system
with one distinguishing direction and the statistical isotropy in the directions perpendicular to the distinguished direction
is reduced to
the two-dimensional (2D) energy-flux vector in the $k_{\perp}$-$k_{\|}$ plane.
The perpendicular and parallel components of the energy-flux vector
by averaging over the azimuthal angles and the signs of $k_z$.
Let us evaluate the perpendicular component of the energy-flux vector going through an arc in a coarse-grained cell represented by the red curve in figure~\ref{fig:3d22d}.
In this example,
the incoming flux $\mathsf{P}_{\perp \mathrm{in}}$ is the sum of the energy fluxes through the sides
inside the arc, $\mathsf{P}_{x\mathrm{in}}^{i-1/2,j,k}$ and $\mathsf{P}_{y\mathrm{in}}^{i,j-1/2,k}$.
Here,
the flux on the cut-cell edge $\mathsf{P}_{x}^{i-1/2,j,k}$,
for example,
is divided
into $\mathsf{P}_{x\mathrm{in}}^{i-1/2,j,k}$ and $\mathsf{P}_{x\mathrm{out}}^{i-1/2,j,k}$
according to the divided lengths.
The outgoing flux $\mathsf{P}_{\perp \mathrm{out}}$ is similarly obtained.
The energy flux through the arc is obtained
by a weighted average of the incoming and outgoing fluxes
as
\begin{eqnarray}
\mathsf{P}_{\perp \mathrm{arc}} =
 \frac{\mathsf{P}_{\perp \mathrm{in}} S_{\mathrm{out}} + \mathsf{P}_{\perp \mathrm{out}} S_{\mathrm{in}}}{S_{\mathrm{in}} + S_{\mathrm{out}}}
 ,
 \label{eq:weightedaverage}
\end{eqnarray}
where $S_{\mathrm{in}}$ and $S_{\mathrm{out}}$ respectively denote
the areas inside and outside the arc,
and $S_{\mathrm{in}} + S_{\mathrm{out}} = \Delta k_x \Delta k_y$.
The perpendicular component of the energy-flux vector is obtained by averaging over these arcs as well as the sign of $k_z$.
The parallel component of the energy-flux vector is obtained
from the $z$ component of the energy flux
by averaging over the azimuthal angles and the signs of $k_z$.

\subsection{Energy-flux vector in homogeneous isotropic turbulence}
\label{sec:check}

To confirm the consistency of the minimal-norm energy-flux vector with the integrated energy fluxes~(\ref{eq:fluxiso}) and (\ref{eq:flux1d}),
the minimal-norm energy-flux vector is obtained
from the DNS of the well-known HIT.
The DNS is performed
with $512^3$ grid points in a cubic box whose volume is $(2\upi)^3$.
The pseudo-spectral method with aliasing removal by the phase shift
is employed to evaluate the nonlinear term,
and hence the maximal wavenumber is approximately $512\sqrt{2}/3\approx 240$.
The Runge-Kutta-Gill method is used for the time integration.
The external force generated by the white noise
is added in the wavenumber space
to the wavenumber mode in $k_{\mathrm{f}}-1/2 \leq |\bm{k}| < k_{\mathrm{f}}+1/2$,
where the forced wavenumber $k_{\mathrm{f}}$ is set to $4$
in this simulation.

In the statistically steady state,
the nonlinear energy transfer rate for each wavenumber mode $T_{\bm{k}}$ is obtained.
The column vector of the energy transfer rates in the coarse-grained cell with the side $\Delta k_x=\Delta k_y=\Delta k_z=\Delta k=3$
is composed,
and the solution vector of the energy flux is obtained as the minimal-norm vector.
The energy-flux vector is converted to the 2D vector
in the $k_{\perp}$-$k_{\|}$ plane
by using the azimuthal average.
The vector in HIT is examined to have only the radial component
in the $k_{\perp}$-$k_{\|}$ plane.
Once the isotropy of the minimal-norm energy-flux vector in HIT
is verified,
the direction of the energy flux in anisotropic turbulence
can be examined by the minimal-norm vector below.

\begin{figure}
 \centerline{
 \includegraphics[scale=.85]{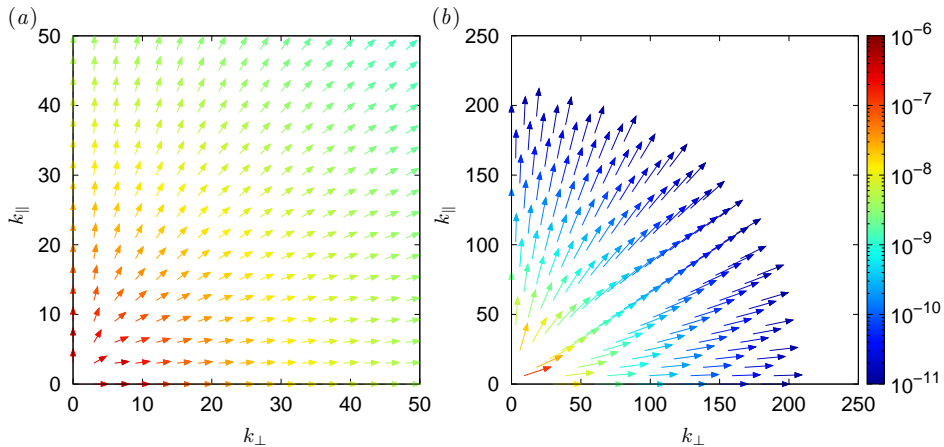}
}
 \caption{
 Energy-flux vector in homogeneous isotropic turbulence
 (\textit{a}) in the small wavenumber range,
 and (\textit{b}) in the whole computational domain.
 }
 \label{fig:fviso}
\end{figure}

The 2D energy-flux vectors in HIT
are drawn in figure~\ref{fig:fviso}.
The 2D energy-flux vectors are obtained from the 3D energy-flux vectors
by averaging over the azimuthal angles and the signs of $k_z$
as written in \S\ref{ssec:vectors}.
Figure~\ref{fig:fviso}(a) is the enlarged view in the small wavenumber range,
while the vectors in a whole computational domain
are drawn
by eliminating some vectors for visibility
in figure~\ref{fig:fviso}(b).
The energy-flux vectors radiate outward at all the wavenumbers
as expected by the forward cascade of energy and its isotropy.
The magnitudes of the energy-flux vectors are large near the origin,
and become small as the magnitudes of the wavenumbers $k=(k_{\perp}^2 + k_{\|}^2)^{1/2}$ become large,
because the areas of the spheres on which the flux is evaluated are proportional to $k^2$
and the energy conservation holds under (\ref{eq:localenergyconservation}).

\begin{figure}
 \centerline{
 \includegraphics[scale=.775]{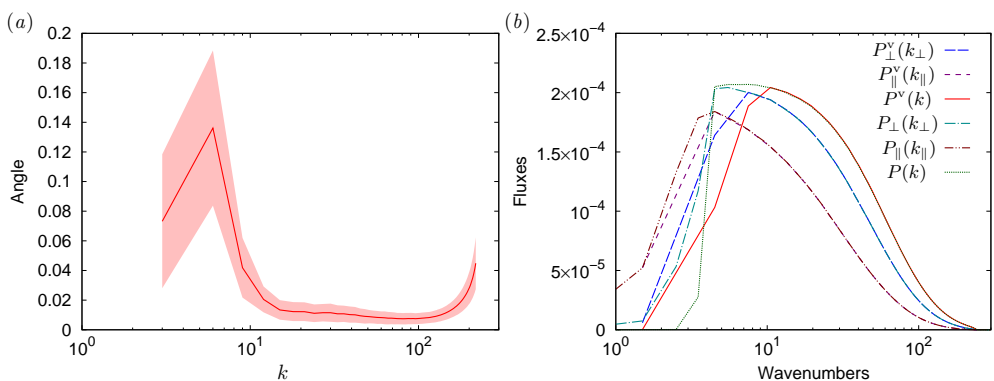}
 }
 \caption{
 (\textit{a}) Angle between $\bm{P}_{\bm{k}}$ and $\bm{k}$ measured in radian.
 The mean and the standard deviation are represented by the solid curve and the filled region, respectively.
  (\textit{b}) Integrated energy fluxes in homogeneous isotropic turbulence.
   The integrated fluxes with the superscript v,
 $P_{\perp}^{\mathrm{v}}$, $P_{\|}^{\mathrm{v}}$, and $P^{\mathrm{v}}$,
 are obtained from the three-dimensional energy-flux vectors,
 while those without the superscript,
 $P_{\perp}$, $P_{\|}$, and $P$,
 are obtained according to (\ref{eq:fluxiso}) and (\ref{eq:flux1d}).
 }
 \label{fig:fv1diso}
  \end{figure}

The direction of the energy flux is quantitatively evaluated
to justify the validity of the minimal-norm energy-flux vector.
The angle between $\bm{P}_{\bm{k}}$ and $\bm{k}$
is defined as
\begin{eqnarray}
\theta_{\bm{k}} = \cos^{-1} \frac{\bm{k} \bcdot \bm{P}_{\bm{k}}}{ |\bm{k}| |\bm{P}_{\bm{k}}|}
,
\end{eqnarray}
and the angle is expected to be $0$ in HIT.
The mean as well as the standard deviation of the angle in the spherical shell $k - \Delta k/2 \leq |\bm{k}| < k+\Delta k/2$
is drawn in figure~\ref{fig:fv1diso}(\textit{a}).
The angles are small,
that is, the energy-flux vectors are radial at all the wavenumbers.
In particular,
the energy-flux vectors in the inertial subrange are almost completely radial.
The small but relatively large angles at the small wavenumbers
are mainly due to the fluctuation of the random external force.
The angles increase near the largest wavenumber
owing to the boundary condition of the energy-flux vectors,
but the magnitudes of the energy-flux vectors are negligibly small.
The isotropy of the energy flux in HIT is successfully validated.

To confirm the consistency of the minimal-norm energy-flux vector with the integrated energy fluxes,
integrated energy fluxes are constituted from the energy-flux vectors.
The integrated perpendicular flux $P_{\perp}^{\mathrm{v}}(k_{\perp})$ is obtained
by the summation of $P_x$ and $P_y$ through the arc (\ref{eq:weightedaverage})
as well as the summation over $k_z$.
The integrated parallel flux $P_{\|}^{\mathrm{v}}(k_{\|})$ is obtained
by the summation of $P_z$ over $k_x$, $k_y$ and the signs of $k_z$.
The isotropic energy flux $P^{\mathrm{v}}(k)$ is obtained
by averaging the energy fluxes
through the inner and outer surfaces of
the cells which cover the sphere with the radius $k$.
These integrated energy fluxes constituted from the energy-flux vectors
are compared with the integrated energy fluxes~(\ref{eq:fluxiso}) and (\ref{eq:flux1d})
in figure~\ref{fig:fv1diso}(\textit{b}).

The perpendicular energy fluxes,
$P_{\perp}^{\mathrm{v}}$ and $P_{\perp}$,
are almost equal to each other.
The difference at the small perpendicular wavenumbers $k_{\perp} < 6$
comes from the discreteness of the coarse-grained cell
during the conversion from $P_x$ and $P_y$ to $P_{\perp}^{\mathrm{v}}$.
The conversion does not make the difference at the large wavenumbers
because the difference between the incoming and outgoing fluxes
due to discreteness of the cells is not so large there.
Moreover,
if the outgoing flux of the cell instead of the weighted average (\ref{eq:weightedaverage}) is used as the flux
to evaluate $P_{\perp}^{\mathrm{v}}$,
then $P_{\perp}^{\mathrm{v}}$ and $P_{\perp}$
are almost equal to each other at the small wavenumbers,
though $P_{\perp}^{\mathrm{v}}$ is then shifted slightly to the smaller wavenumber at the large wavenumbers.
Similarly,
the radial energy fluxes, $P^{\mathrm{v}}$ and $P$, are almost equal to each other,
though the difference at the small wavenumbers also emerges owing to the discreteness of the cells.
Because $P_{\|}^{\mathrm{v}}$ is evaluated exactly on the cell faces,
and is not affected by the discreteness of the cells,
$P_{\|}^{\mathrm{v}}$ and $P_{\|}$ are equal to each other all over the wavenumbers
within the convergence criterion during the calculation by the GMRES method.
Note that $P_{\|}^{\mathrm{v}}$ has its value at $k_{\| i} = 3(i-1/2)$
owing to the coarse-graining with $\Delta k=3$
as written above,
while $P_{\|}$ does at $k_{\| j} = j-1/2$,
where $i$ and $j$ here denote positive integers.
Therefore,
these energy fluxes obtained from the energy-flux vectors
are equal to the integrated energy fluxes
except for the difference due to the discreteness of the cells.
In this way,
the magnitudes of the energy-flux vectors
obtained by the Moore--Penrose inverse
are quantitatively consistent with the integrated energy fluxes
in HIT.

\section{Application to strongly rotating turbulence}
\label{sec:application}

In this section,
the energy flux as the minimal-norm solution of the continuity equation of energy
is examined in rotating turbulence
by comparing the vector with the theoretical predictions.
Direct numerical simulations of rotating turbulence are performed
according to (\ref{eq:governingeqrtw})
by using the pseudo-spectral method.
In the following simulations of the rotating turbulence,
the periodic box has dimensions of $2\upi \times 2\upi \times 8\upi$
so that $k_x, k_y \in \mathbb{Z}$ and $k_z \in \mathbb{Z}/4$.
Here,
the periodic box long in the $z$ direction
is used because of the anisotropy at the large scales.
The non-dimensional numbers which characterise the rotating turbulence
are the turbulent Reynolds number
$\Rey_{\mathrm{t}} = \overline{\varepsilon}^{1/3}/(\nu k_{\mathrm{f}}^{4/3}) = (k_{\mathrm{\eta}}/k_{\mathrm{f}})^{4/3}$
and the rotational Reynolds number
$\Rey_{\Omega} = \overline{\varepsilon}/(\nu \Omega^2) = (k_{\mathrm{\eta}}/k_{\Omega})^{4/3}$,
where $\overline{\varepsilon}$ is the energy dissipation rate
and $k_{\mathrm{\eta}}$ is the Kolmogorov wavenumber.

\subsection{Theoretical prediction of energy flux in rotating turbulence}
\label{sec:WTTandEnergyFlux}

When the nonlinear term, the viscosity and the external force in the governing equation~(\ref{eq:governingeqrtw})
are neglected,
the linear inviscid equation can be written as
\begin{equation}
\partial a_{\bm{k}}^{s_{\bm{k}}} / \partial t =
-\mathrm{i} s_{\bm{k}} \sigma_{\bm{k}} a_{\bm{k}}^{s_{\bm{k}}}
,
\label{eq:gov-eq4ca}
\end{equation}
where the complex amplitude $a_{\bm{k}}^{s_{\bm{k}}}$ is defined as
$a_{\bm{k}}^{s_{\bm{k}}} = \bm{u}_{\bm{k}} \bcdot \bm{h}_{\bm{k}}^{-s_{\bm{k}}}$
according to the helical-mode decomposition~\citep{alexakis_2017,PhysRevE.68.015301,doi:10.1063/1.870022,:/content/aip/journal/pofa/5/3/10.1063/1.858651},
and $s_{\bm{k}}=\pm 1$ denotes the sign of the helicity of the inertial wave.
The basis is expressed as $\bm{h}_{\bm{k}}^{s_{\bm{k}}} = (\bm{e}_1 +\mathrm{i} s_{\bm{k}} \bm{e}_2)/\sqrt{2}$,
where
$(\bm{e}_1, \bm{e}_2)
= ( \bm{e}_z \times \bm{k} / |\bm{e}_z \times \bm{k}|,
\bm{k} \times (\bm{e}_z \times \bm{k}) / |\bm{k} \times (\bm{e}_z \times \bm{k})|)$
for $k_{\perp} \neq 0$,
and $(\bm{e}_1,\bm{e}_2) = (\bm{e}_x, \bm{e}_y)$ for $k_{\perp} = 0$.
The linear dispersion relation is given by $\sigma_{\bm{k}} = 2\Omega k_z/k$.
The linear inviscid equation (\ref{eq:gov-eq4ca}) has the wave solutions
$\bm{u}(\bm{x}) \propto \bm{h}_{\bm{k}}^{s_{\bm{k}}} \mathrm{e}^{\mathrm{i} (\bm{k}\bcdot\bm{x} - s_{\bm{k}} \sigma_{\bm{k}} t)} + \mathrm{c.c.}$
called inertial waves.

In the wave-dominant range,
the period of the inertial wave is considered to be much shorter
than the eddy turnover time,
that is, $1/\sigma_{\bm{k}} \ll (k^2 \overline{\varepsilon})^{-1/3}$~\citep{:/content/aip/journal/pof2/26/3/10.1063/1.4868280}.
On the premise of the local nonlinear interaction,
\citet{PhysRevE.68.015301} applied the weak turbulence theory
to the inertial waves in the strongly rotating turbulence,
and he found that
only a small energy transfer along $\bm{\Omega}$ is allowed
by the resonance condition:
\begin{equation}
 \bm{k} = \bm{k}_1 + \bm{k}_2,
\quad
 s_{\bm{k}} \sigma_{\bm{k}} = s_{\bm{k}_1} \sigma_{\bm{k}_1} + s_{\bm{k}_2} \sigma_{\bm{k}_2}
 .
\end{equation}
He also discussed
the nonlocality of the nonlinear interactions,
which generates the anisotropy.
\citet{:/content/aip/journal/pofa/5/3/10.1063/1.858651} applied his idea of
the instability assumption on the nonlinear energy transfers
to predict the anisotropic energy transfer among the wavenumber modes
and obtained similar results.
The energy flux parallel to the perpendicular wavenumber axis is theoretically expected
in the wave-dominant range.

At the wavenumbers larger than the Zeman wavenumber,
where the Coriolis period is larger than the eddy turnover time,
the rotation is negligible at such large wavenumbers.
Thus,
the isotropic Kolmogorov turbulence appears at the larger wavenumbers,
and the energy flux is isotropic like that in HIT shown in figure~\ref{fig:fviso}.

The energy flux in the buffer range
between in the weak turbulence range and the isotropic Kolmogorov turbulence range
is expected
to connect the energy flux parallel to the perpendicular wavenumber axis
in the inertial-wave turbulence
and the isotropic energy flux in the isotropic Kolmogorov turbulence.
\citet{nazarenko3488critical}
predicted that the energy is transferred along the wavenumbers
at which the period of the inertial wave is comparable with the eddy turnover time,
after the energy is carried to such wavenumbers
by the resonant interactions among inertial waves.
That is,
the critical balance predicts that
the energy is transferred to large $k_{\|}$'s in the buffer range
in figure~\ref{fig:coexistrt}.

\subsection{Minimal-norm energy-flux vector in rotating turbulence}

\subsubsection{Comparison with the weak turbulence theory}
\label{sec:ComparisonWithWTT}

In order to validate the minimal-norm energy-flux vector in anisotropic turbulence,
it is compared with the energy flux in the weak turbulence theory.
A direct numerical simulation of rotating turbulence
where the external force is applied to the wavenumber modes $k_{\perp} \approx 0$
is performed
to compare the energy-flux vectors
directly with the prediction of the weak turbulence theory of the inertial waves.
Here,
the random external force is applied to the small perpendicular wavenumber modes
$k_x, k_y= 0, \pm 1$ and $|k_z| \leq 50$
in a DNS with $256\times 256\times 1024$ grid points.
In this simulation,
the rotational Reynolds number is evaluated as $\Rey_{\Omega} \approx 0.4$.
Although the turbulent Reynolds number is not well defined in this simulation
because the forced wavenumbers are widely distributed,
the turbulent Reynolds number is considered to be small to investigate the energy flux in the weak turbulence.
Time averaging to obtain the energy transfer rate $T_{\bm{k}}$
required in the local energy conservation~(\ref{eq:localenergyconservation})
as well as the energy spectra
is performed
in the statistically steady state.

\begin{figure}
 \centerline{
 \includegraphics[scale=.65]{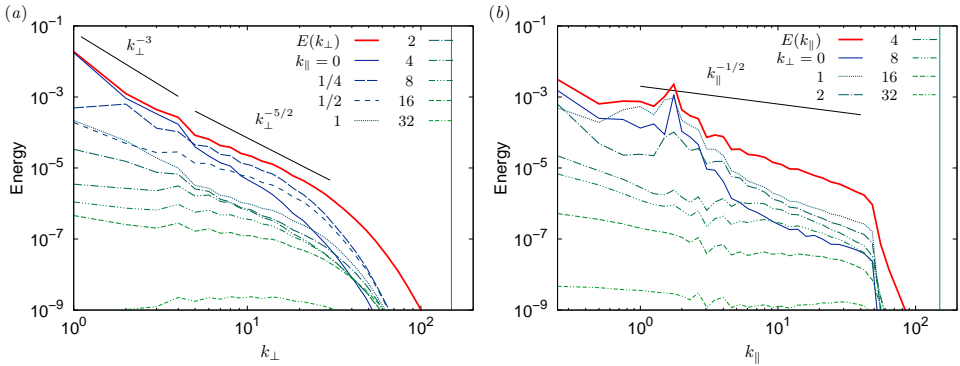}
 }
 \caption{
 Kinetic energy spectra
 when the external force is applied to the small perpendicular wavenumbers
 (\textit{a}) for each $k_{\|}$ as function of $k_{\perp}$
 and
 (\textit{b}) for each $k_{\perp}$ as function of $k_{\|}$.
 }
 \label{fig:spslicewt}
\end{figure}

In the weak turbulence theory,
the energy spectrum of the inertial waves
is predicted as
$E(k_{\perp}, k_{\|}) \propto k_{\perp}^{-5/2} k_{\|}^{-1/2}$~\citep{PhysRevE.68.015301}.
To observe the anisotropic spectra,
the kinetic energy spectra
for each $k_{\|}$ as function of $k_{\perp}$
 and
 for each $k_{\perp}$ as function of $k_{\|}$
 are drawn in figure~\ref{fig:spslicewt}.
 The energy spectrum for each $k_{\|}$ as a function of $k_{\perp}$,
 for example,
 is defined as
\begin{eqnarray}
 E_{k_{\|}}(k_{\perp}) =
\frac{1}{\Delta k_{\perp}}
 {\sum_{\bm{k}_{\perp}^{\prime}}}^{\prime}
 \frac{1}{\Delta k_{\|}}
 {\sum_{k_z^{\prime}}}^{\prime}
 \frac{1}{2} \langle |\bm{u}_{\bm{k}_{\perp}^{\prime}, k_z^{\prime}}|^2\rangle
,
\end{eqnarray}
where the summations
${\sum_{\bm{k}_{\perp}^{\prime}}}^{\prime}$ and ${\sum_{k_{\|}^{\prime}}}^{\prime}$
are respectively taken over
$k_{\perp} - \Delta k_{\perp}/2 \leq |\bm{k}_{\perp}^{\prime}| < k_{\perp} + \Delta k_{\perp}/2$
and $k_{\|} -\Delta k_{\|}/2 \leq |k_z^{\prime}| < k_{\|} + \Delta k_{\|}/2$,
and $\Delta k_{\perp}$ and $\Delta k_{\|}$ are the bin widths to obtain the spectrum.
The corresponding integrated spectra
as a function of $k_{\perp}$
\begin{eqnarray}
E(k_{\perp})
 =
\frac{1}{\Delta k_{\perp}}
 {\sum_{\bm{k}_{\perp}^{\prime}}}^{\prime}
 {\sum_{k_z^{\prime}}}
 \frac{1}{2} \langle |\bm{u}_{\bm{k}_{\perp}^{\prime}, k_z^{\prime}}|^2\rangle
 = \int \mathrm{d}k_{\|} E_{k_{\|}}(k_{\perp})
\end{eqnarray}
are also drawn in figure~\ref{fig:spslicewt}.

The perpendicular-wavenumber spectra of the energy are
close to $k_{\perp}^{-5/2}$
in the wavenumber range of $4 \lessapprox k_{\perp} \lessapprox 30$ and $10 \lessapprox k_{\|} \lessapprox 30$.
Similarly,
the parallel-wavenumber spectra of the energy are close to $k_{\|}^{-1/2}$
in the same wavenumber range.
Moreover,
the parallel-wavenumber spectra have abrupt drops at $k_{\|} \approx 50$,
and
the energy injected by the external force is rarely transferred
to the large parallel wavenumbers $k_{\|} > 50$.
It is consistent with the prediction of the weak turbulence theory of the inertial waves,
in which the resonant interactions transfer the energy
only to the wavenumbers having the same $k_{\|}$.
These spectra demonstrate that
the inertial-wave turbulence appears
in the wavenumber range of $4 \lessapprox k_{\perp} \lessapprox 30$ and $5 \lessapprox k_{\|} \lessapprox 30$.
The energy spectra show
that this DNS is appropriate to compare the energy-flux vector
with the prediction of the weak turbulence theory.

\begin{figure}
 \centerline{
 \includegraphics[scale=.85]{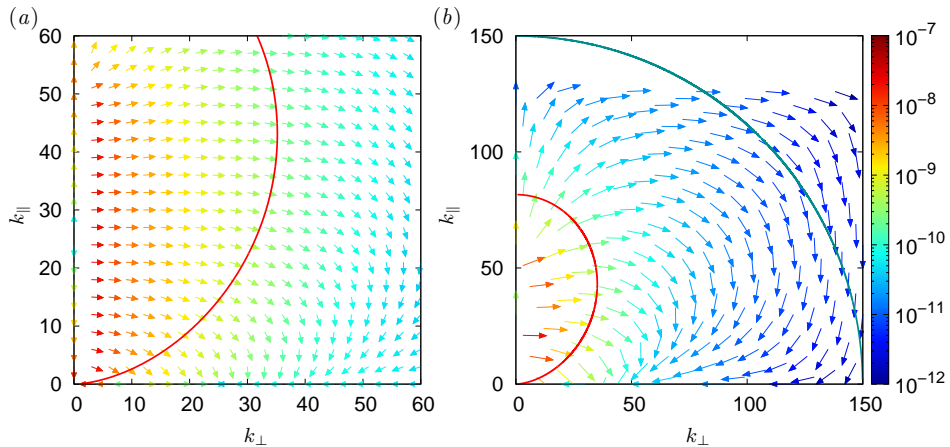}
 }
 \caption{
 Energy-flux vectors
 when the external force is applied to the small perpendicular wavenumbers
 (\textit{a}) in the small wavenumber range,
 and (\textit{b}) in the whole computational domain.
 The critical wavenumber and the Zeman wavenumber
 are respectively drawn in the red and cyan curves.
 }
 \label{fig:fvwt}
\end{figure}

The energy-flux vector in the rotating turbulence
whose spectra are drawn in figure~\ref{fig:spslicewt}
is obtained by the procedure written in \S\ref{ssec:vectors}.
The perpendicular and parallel components of the energy-flux vectors
are respectively obtained from the $x$ and $y$ components and the $z$ components
by averaging over the azimuthal angles and the signs of $k_z$.
The energy-flux vectors in the $k_{\perp}$-$k_{\|}$ plane are drawn
in figure~\ref{fig:fvwt}.

The weak-wave turbulence is expected to exist
at the wavenumbers,
where the linear wave period $\tau_{\mathrm{w}} = (2\Omega k_{\|}/k)^{-1}$
is much shorter than the eddy turnover time $\tau_{\mathrm{e}} = (k^2 \overline{\varepsilon})^{-1/3}$.
The critical wavenumber is evaluated to appear
at $\tau_{\mathrm{w}} = \tau_{\mathrm{e}}/3$
in magnetohydrodynamic turbulence~\citep{PhysRevLett.116.105002}
and stratified turbulence~\citep{PhysRevFluids.4.104602}.
On the other hand,
the Coriolis force affects little
and the isotropic Kolmogorov turbulence appears
at the wavenumber range
where the Coriolis period $\tau_{\Omega} = (2\Omega)^{-1}$ is larger than the eddy turnover time $\tau_{\mathrm{e}}$.
The separation wavenumber of the isotropic Kolmogorov turbulence
is known as the Zeman wavenumber $k_{\Omega}$.
The buffer range should exist between the critical wavenumber and the Zeman wavenumber.
The curves which represent the critical wavenumber (red) and the Zeman wavenumber (cyan)
are drawn in figure~\ref{fig:fvwt}.

The energy-flux vectors at the wavenumber modes $k_{\|} \leq 50$ in the weak turbulence range
are almost completely parallel to the $k_{\perp}$ axis,
and
the energy provided by the external force rarely goes to the modes $k_{\|} > 50$.
Thus,
the wavenumber modes at $k_{\|} > 50$ in the weak turbulence range have little energy.
The minimal-norm vector
can well reproduce the anisotropic energy flux
in accordance with the resonant interactions among the inertial waves
locally in the wavenumber space.

The energy flux in the range
where $\tau_{\mathrm{w}} < \tau_{\mathrm{e}}/3$
and the weak-wave turbulence is expected to exist
does not necessarily demonstrate the perpendicular flux.
It results from the fact that
the wavenumber modes at $k_{\|}>50$ have little energy
and the modes in the range are subordinate to the modes at $k_{\|}<50$
having much larger energy.

The energy-flux vectors turn the direction
near the wavenumber modes having $\tau_{\mathrm{w}} = \tau_{\mathrm{e}}/3$.
In this case,
the energy flux along the critical wavenumbers
that gives the isotropisation of energy in the buffer range
is not observed.
In fact,
the energy transferred via the resonant interactions
moves on to the 2D modes $k_{\|} = 0$.

\begin{figure}
 \centerline{
 \includegraphics[scale=1.]{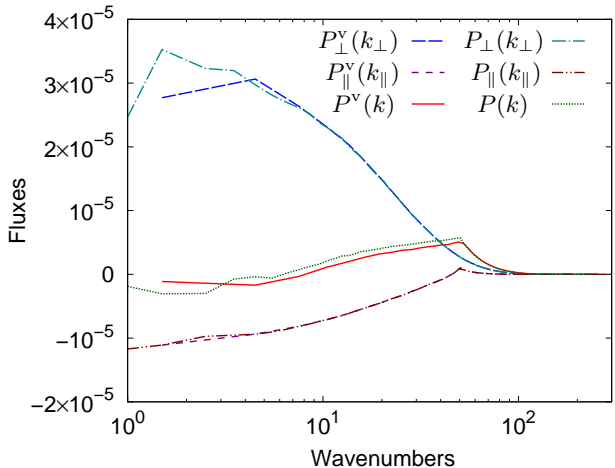}
 }
 \caption{
 Perpendicular, parallel and isotropic integrated fluxes
 when the external force is applied to the small perpendicular wavenumbers.
 The superscript v denotes the integrated fluxes
 obtained from the three-dimensional energy-flux vectors.
 See also caption for figure~\ref{fig:fv1diso}.
 }
 \label{fig:fv1dwt}
\end{figure}

The integrated energy fluxes obtained from the 3D energy-flux vectors
in the simulation where the external force is applied to the small perpendicular wavenumbers
are almost equal to the corresponding integrated energy fluxes
as drawn in figure~\ref{fig:fv1dwt}.
The difference between $P_{\perp}^{\mathrm{v}}$ and $P_{\perp}$
at the small perpendicular wavenumbers $k_{\perp} < 6$
comes from the discreteness of the coarse-grained cell
during the conversion from $P_x$ and $P_y$ to $P_{\perp}^{\mathrm{v}}$
as seen in the isotropic turbulence (figure~\ref{fig:fv1diso}).
The difference between $P^{\mathrm{v}}$ and $P$
at $k < 50$ where the external force directly affects
comes from the discreteness of the coarse-grained cell
during the averaging the energy fluxes
over the azimuthal angles
or through the inner and outer surfaces of the cells.
The negative parallel flux, $P_{\|}<0$,
appearing at $k_{\|}<50$ where the external force affects
shows that most of the energy provided by the external force
goes to the small parallel wavenumbers,
and is dissipated there.

\subsubsection{Energy flux in direct numerical simulation with isotropic forcing}
\label{sec:EnergyFluxLargeScaleForcing}

Another DNS
where more commonly-used external force is employed
is performed with $512 \times 512 \times 2048$ grid points.
One may think that the resolution is not so high as that in recent high-resolution simulations.
However, such high-resolution simulation is not easily performed
because a long-time integration is required to obtain the weak inertial-wave turbulence
where the energy is transferred by the resonant interactions.

The 3D three-component random force
is added isotropically to the small wavenumbers in $k_{\mathrm{f}}-1/2 \leq |\bm{k}| < k_{\mathrm{f}}+1/2$,
where the forced wavenumber $k_{\mathrm{f}}$ is set to $4$.
In this DNS,
the turbulent Reynolds number and the rotational Reynolds number
are respectively evaluated
as $\Rey_{\mathrm{t}} \approx 1.6\times 10^2$ and $\Rey_{\Omega} \approx 2.8$.

\begin{figure}
\centerline{
 \includegraphics[scale=1]{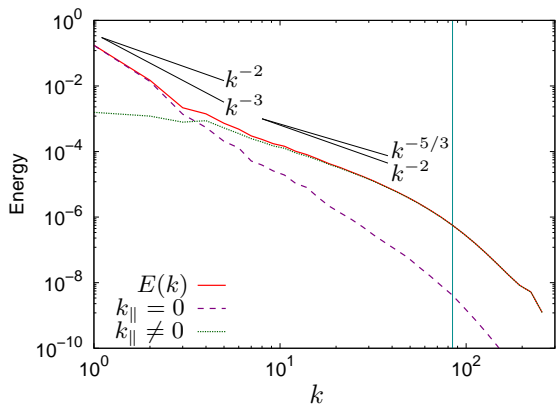}
}
 \caption{
 Energy spectrum,
 that given only by two-dimensional modes ($k_{\|}=0$),
 and that given by three-dimensional modes ($k_{\|} \neq 0$).
 The vertical line indicates the Zeman wavenumber.
}
 \label{fig:sprot}
\end{figure}

The energy spectrum as a function of the norms of wavenumbers
obtained in the DNS
is drawn in figure~\ref{fig:sprot}.
The energy spectrum at the wavenumbers smaller than those of the external force $k_{\mathrm{f}} \approx 4$
is as steep as $k^{-3}$.
The 2D flows,
which are uniform in the direction parallel to the rotational axis, i.e., $k_{\|}=0$,
account for a large fraction of the energy at the small wavenumbers,
though the number of the 2D modes is much smaller than
that of the 3D modes which have $k_{\|} \neq 0$.
In fact,
a large-scale columnar vortex is formed.
Note that a statistically steady state can be achieved without small-wavenumber drag~\citep{PhysRevE.85.036315}.
At the wavenumbers larger than $k_{\mathrm{f}}$,
the energy spectrum is dominated by the 3D flows,
and is less steep than at the small wavenumbers.
In this energy spectrum,
the isotropic Kolmogorov turbulence cannot be clearly observed,
because the Zeman wavenumber is close to the Kolmogorov wavenumber
as well as the cutoff wavenumber.

\begin{figure}
\centerline{
 \includegraphics[scale=.65]{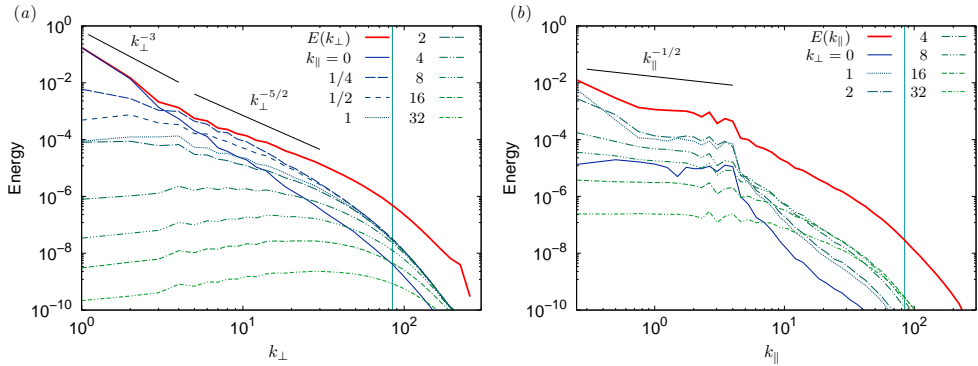}
}
 \caption{
 Kinetic energy spectra
 (\textit{a}) for each $k_{\|}$ as function of $k_{\perp}$
 and
 (\textit{b}) for each $k_{\perp}$ as function of $k_{\|}$.
 }
 \label{fig:spslice}
\end{figure}

The energy spectra are drawn in figure~\ref{fig:spslice}.
The spectra are close to the prediction of the weak turbulence theory,
$E(k_{\perp}, k_{\|}) \propto k_{\perp}^{-5/2} k_{\|}^{-1/2}$,
in the wavenumber range of $4 \lessapprox k_{\perp} \lessapprox 30$ and $1/4 \lessapprox k_{\|} \lessapprox 4$.
The abrupt drops at $k_{\|} \approx 4=k_{\mathrm{f}}$ in the parallel-wavenumber spectra
indicates that little energy is transferred
to the large parallel wavenumbers $k_{\|} > k_{\mathrm{f}}$
consistently with the weak turbulence theory.
These spectra demonstrate that
the inertial-wave turbulence appears
in the wavenumber range of $4 \lessapprox k_{\perp} \lessapprox 30$ and $1/4 \lessapprox k_{\|} \lessapprox 4$.

\begin{figure}
 \centerline{
 \includegraphics[scale=.85]{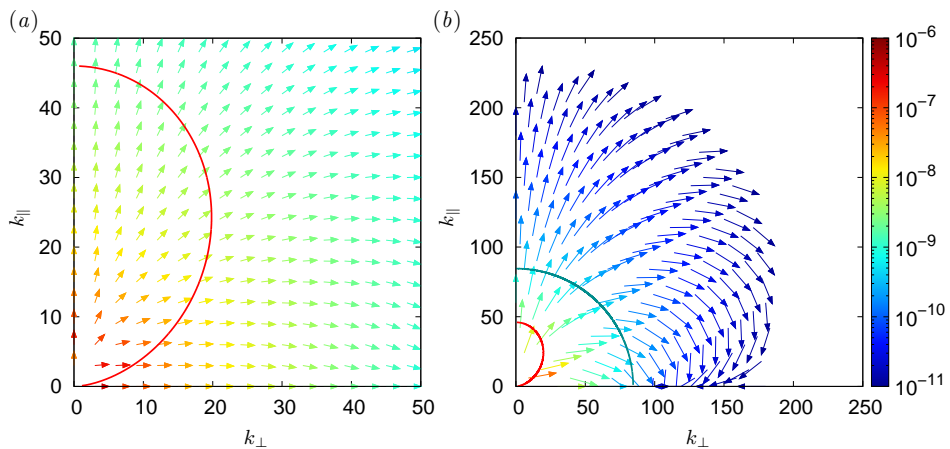}
}
 \caption{
 Energy-flux vectors
where the isotropic external force is applied
 (\textit{a}) in the small wavenumber range,
 and (\textit{b}) in the whole computational domain.
 See also the caption for figure~\ref{fig:fvwt}.
 }
 \label{fig:fvrot}
\end{figure}

The energy-flux vectors in the $k_{\perp}$-$k_{\|}$ plane
in the isotropically forced rotating turbulence
are drawn in figure~\ref{fig:fvrot}.
The energy-flux vectors roughly in the range
of $k_{\perp} \lessapprox 30$ and $k_{\|} \lessapprox 4$
are almost parallel to the $k_{\perp}$ axis.
The perpendicular flux is consistent with the resonant interactions of the inertial waves
in the weak turbulence theory.
In fact,
the range where the perpendicular flux is observed
corresponds to the range where the weak inertial-wave spectra are observed in figure~\ref{fig:spslice}.

On the other hand,
the vector field of the energy flux
has a sink at $(k_{\perp}, k_{\|}) \approx (100, 0)$.
In this simulation,
the energy flux along the critical wavenumbers
predicted by the critical balance is not observed.
This discrepancy is discussed in the next section.

\begin{figure}
 \centerline{
 \includegraphics[scale=1.]{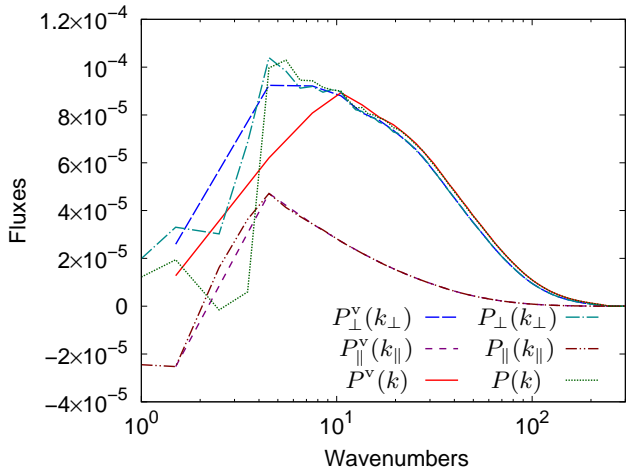}
 }
 \caption{
 Perpendicular, parallel and isotropic integrated fluxes in isotropically forced rotating turbulence.
 The superscript v denotes the integrated fluxes
 obtained from the three-dimensional energy-flux vectors.
 See also caption for figure~\ref{fig:fv1diso}.
 }
 \label{fig:comp1d}
\end{figure}

To confirm the consistency of the energy-flux vectors with the commonly-used integrated energy fluxes,
the integrated fluxes are obtained from the 3D flux vectors
and compared with those obtained by (\ref{eq:fluxiso}) and (\ref{eq:flux1d})
in figure~\ref{fig:comp1d}.
The isotropic and perpendicular fluxes obtained from the 3D flux vectors
are almost equal to those obtained by (\ref{eq:fluxiso}) and (\ref{eq:flux1d})
except for the small wavenumbers $k, k_{\perp} < 8$.
The difference at the small wavenumbers comes from the discreteness of the cells
and the averaging for $k_{\perp}$ and $k$.
The parallel flux obtained from the 3D flux vectors
and the corresponding integrated flux
are the same within the convergence criterion adopted in the GMRES method.
These agreement and disagreement have been seen in the energy flux
in HIT (figure~\ref{fig:fv1diso}(\textit{b})).
Therefore,
the energy-flux vectors are consistent with the integrated energy fluxes,
which have been usually used in the anisotropic turbulence.

\section{Discussion}
\label{sec:discussion}

\begin{figure}
 \centerline{
 \includegraphics[scale=1]{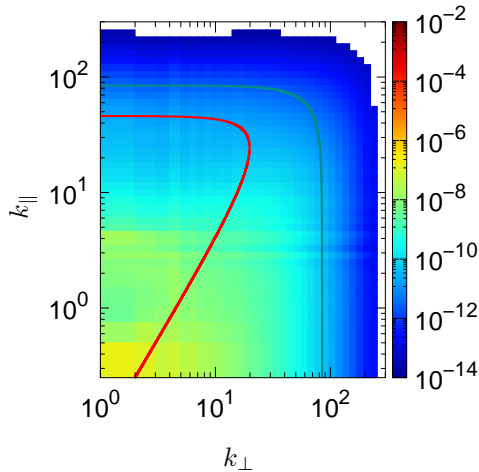}
 }
 \caption{
 Two-dimensional spectrum of energy dissipation rate in rotating turbulence.}
 \label{fig:dissipationrot}
\end{figure}

The minimal-norm energy-flux vector successfully shows the local structures of the energy flux
both in HIT corresponding to the strong turbulence (figure~\ref{fig:fviso})
and in the anisotropic weak-wave turbulence (figure~\ref{fig:fvwt}).
Thus,
one can expect that
the minimal-norm vector can be applied also in the buffer range
between the weak turbulence range and the strong turbulence range.
Contrasting to the prediction by the critical balance, however,
the energy flux along the wavenumbers
at which the linear wave periods and the nonlinear eddy turnover times are comparable
is not observed in the present simulations.
In fact,
the energy flux appears to have a sink in the buffer range.

It is important to clarify the reason for the discrepancy
between the numerical result shown in the present paper
and the prediction of the weak turbulence theory.
To find the reason,
the 2D spectrum of energy dissipation rate
in the DNS with the isotropic external force
is drawn in figure~\ref{fig:dissipationrot}.
Note again that the dissipation rate and the energy transfer rate statistically balance at all the wavenumbers except for the forced wavenumbers
in the ideal statistically steady states.
The 2D spectrum
is defined as
\begin{equation}
 D(k_{\perp}, k_{\|})
 =
 \frac{1}{2\upi k_{\perp}}
 \frac{1}{\Delta k_{\perp}}
 {\sum_{\bm{k}_{\perp}^{\prime}}}^{\prime}
 \frac{1}{\Delta k_{\|}}
 {\sum_{k_z^{\prime}}}^{\prime}
 \nu (|\bm{k}_{\perp}^{\prime}|^2 + k_{z}^{\prime 2}) \langle |\bm{u}_{\bm{k}_{\perp}^{\prime}, k_z^{\prime}}|^2\rangle
 ,
\end{equation}
where $1/(2\upi k_{\perp})$ is introduced to easily compare the 2D spectrum with the isotropic one.
The dissipation rate
is large at the forced wavenumbers
and at the wavenumbers of the large-scale columnar vortex which has $k_{\|} \approx 0$.
In addition,
the dissipation is remarkable
in the buffer range between the wave-dominant range surrounded by the critical wavenumber
and the range of the isotropic turbulence where the wavenumber is larger than
the Zeman wavenumber.

The energy provided by the external force is transferred to this buffer range
by the resonant interactions as seen in figure~\ref{fig:fvrot}.
The critical balance predicts
that the energy which reaches the buffer range must start to be isotropised
by changing the direction of transfer
in this range.
However,
the sink due to the strong dissipation is observed in this range
instead of the isotropisation.
In this sense,
the dissipation plays a dominant role in the buffer range in the present simulations,
and the large dissipation there is not considered in the theory of the critical balance.
A large range of the isotropic turbulence at the large wavenumbers
is required to determine whether or not the energy is transferred along the critical wavenumbers.
The wave-dominant range cannot be small
because the resonant interactions are relatively sparse in the wavenumber space.
Then,
the high-resolution simulations
where both of the wavenumber ranges of the weak turbulence and the isotropic turbulence are large enough
are required.

Let us consider the asymptotic behaviour of the large dissipation
which appears in the buffer range in the present simulations
at the large Reynolds numbers.
Suppose that the higher-resolution simulation
has a smaller kinematic viscosity
without changing the magnitude of the external force
and that of the anisotropy, i.e., the system's rotation rate.
The buffer range in the higher-resolution simulation remains at almost the same wavenumber range
as that in the lower-resolution simulation,
because the linear time scales are unchanged
and the total energy dissipation rate and hence eddy turnover time
are almost the same.
On the other hand,
the relatively large dissipation in the buffer range should become smaller.
Then,
some of the energy pass through the dissipation in the buffer range,
and the energy flux along the critical wavenumbers that brings the isotropisation could be possible.

The critical balance obviously assumes that
the dissipation affects only at the large wavenumbers outside of the isotropic Kolmogorov turbulence range.
In fact, however, the dissipation is not localised;
the dissipation spectrum in the inertial subrange of the isotropic Kolmogorov turbulence
slowly increases as $k^{1/3}$, for example.
Thus,
much higher-resolution simulation is required
to realise the critical balance.
Note that
the large-eddy simulations and the hyper-viscosity should be carefully used
because the change of the dissipation can alter the direction of the energy flux.

The critical balance predicts the isotropisation
in the buffer range between in the weak turbulence range and the isotropic Kolmogorov turbulence range.
Such ranges are characterised by the time scales of dominant physical mechanisms:
the linear wave period, the characteristic time of the anisotropic system and the eddy turnover time.
When the external force is applied to the small perpendicular wavenumbers,
$\tau_{\mathrm{w}} < \tau_{\mathrm{e}}/3$ in the inertial-wave turbulence range
as shown in figures~\ref{fig:spslicewt} and \ref{fig:fvwt}.
On the other hand,
when the external force is applied isotropically to the small wavenumbers,
the inertial-wave turbulence range goes into the range where $\tau_{\mathrm{w}} \sim \tau_{\mathrm{e}}$
as shown in figures~\ref{fig:spslice} and \ref{fig:fvrot}.
It is mainly due to the high directionality
of the energy transfer in the inertial-wave turbulence.
The wavenumber range of the inertial-wave turbulence
depends on the wavenumbers at which the external force is applied
as pointed out by \citet{nazarenko3488critical}.
The misalignment of the inertial-wave turbulence
might indicate the existence of key mechanisms other than the time scales
in the energy flux in the anisotropic turbulence.

When the energy due to the external force is injected
at the middle wavenumbers,
the energy is expected to cascade
to the smaller wavenumbers as 2D turbulence
and to the larger wavenumbers as 3D turbulence~\citep{ALEXAKIS20181}.
If the high-resolution simulations could be performed,
such split energy cascades should be observed.
In fact,
the energy flux toward small $k_{\perp}$
at $k_{\perp} < 10$ and $k_{\|}\approx 0$
can be observed in figure~\ref{fig:fvwt}(\textit{a}).
This 2D inverse cascade suggests that
the split energy cascade can be observed
by using the energy-flux vectors proposed here.

The two ansatzes,
the net locality of the energy balance
and the efficiency of the energy transfer,
are introduced to obtain the minimal-norm energy-flux vector.
It is assumed by the former that
the local energy transfer is dominant in the wavenumber space,
and the local energy conservation holds.
Such locality is essential
when the detailed local structures of the energy flux is discussed.
It is assumed by the latter that
the minimal-norm vector among other possible solutions of the local energy conservation
is irrotational and excludes local circulations of the energy transfer.
This efficiency of the energy transfer is natural
because the local circulations make closed loops of the energy fluxes,
and do not intrinsically contribute the net energy flux.
Although these ansatzes are introduced,
the energy-flux vectors are consistent with the integrated energy fluxes,
which have been conventionally used.

\section{Summary}

In anisotropic turbulence,
the energy flux in the wavenumber space should be considered
as a vector at each wavenumber mode,
because different kinds of turbulence coexist inhomogeneously at each scale.
The integrated energy fluxes which have been used
to exhibit the anisotropy in the energy transfer
cannot reveal detailed local structures of the energy flux,
because these integrated fluxes inevitably consist of multiple turbulence ranges.
In this work,
the minimal-norm energy-flux vector
obtained by using the Moore--Penrose inverse
was proposed to uniquely determine the energy-flux vectors
based on the two ansatzes:
the net locality of the energy balance
and the efficiency of the energy transfer.
The latter is equivalent to the irrotationality of the energy-flux vectors.

The minimal-norm energy-flux vector is tested in
both strongly nonlinear isotropic turbulence and anisotropic weak-wave turbulence.
The minimal-norm vectors successfully show
the isotropic energy flux that radiates outward
in homogeneous isotropic turbulence.
In the rotating turbulence,
the minimal-norm flux in the weak turbulence range demonstrates
the energy flux parallel to the perpendicular wavenumber axis.
Namely,
the minimal-norm vectors
successfully display the energy flux consistent
with the energy transfer due to the resonant interactions among the inertial waves
according to the weak turbulence theory.

On the other hand,
the energy flux along the critical wavenumbers
predicted by the critical balance
was not observed
 in the buffer range between in the weak turbulence range and the isotropic Kolmogorov turbulence range
 in the rotating turbulence.
The inconsistency of the energy-flux vector proposed here
with the critical balance
does not indicate the failure of the proposed method
but results from the appearance of the large dissipation
in the buffer range.
This inconsistency will be resolved
when much higher-resolution simulations are performed.
In addition,
a two-dimensional turbulence such as the quasi-geostrophic turbulence is more suitable
to obtain the coexistence of the large wavenumber ranges of
the weak-wave turbulence and the strong turbulence
as well as the buffer range.
The energy-flux vector in the two-dimensional turbulence
will be reported elsewhere.

\begin{acknowledgments}
Numerical computation in this work was carried out
 at the Yukawa Institute Computer Facility, Kyoto University
 and Research Institute for Information Technology, Kyushu University.
This work was partially supported by JSPS KAKENHI Grant
No.~15K17971, No.~16K05490, No.~17H02860, No.~18K03927, and No.~19K03677.
\end{acknowledgments}

%\section*{Declaration of Interests}
The authors report no conflict of interest.

\bibliographystyle{jfm}

\end{document}